# Analysis of ellipsometric data obtained from curved surfaces

J. Křepelka

*Joint Laboratory of Optics of Palacky University and Institute of Physics of Academy of Sciences of the Czech Republic, 17. listopadu 50a, 772 07 Olomouc, Czech Republic*

**Abstract:** This article deals with quantitative error analysis resulting from ellipsometric data obtained from measurement on curved surfaces including the influence of non-collimated beams. Numerical model based on the combination of geometrical and wave optics is restricted to the example of single dielectric layer deposited on the substrate with complex index of refraction. Three methods for averaging measurable ellipsometric data are compared.

## 1. Normal to the curved surface

Let $z = z(x, y)$ is a function describing the curved surface in a Cartesian co-ordinate system, then a vector perpendicular to the tangential plane in the point $\boldsymbol{r}(x, y) = (x, y, z(x, y))$ of the surface is $\boldsymbol{u}(x, y) = -\mathrm{grad}(z(x, y) - z)$. Hence the unit normal vector is $\boldsymbol{n}(x, y) = \dfrac{\boldsymbol{u}(x, y)}{u(x, y)}$, where $u(x, y) = \|\boldsymbol{u}(x, y)\|$ is a size (norm) of the vector $\boldsymbol{u}(x, y)$. Especially for the upper half-sphere $z(x, y) = \sqrt{R^2 - x^2 - y^2}$, $x^2 + y^2 \leq R^2$, where $R$ is a sphere radius (and centre of the sphere coincides with the origin of co-ordinate system), we have a unit vector normal to the surface at the point $\boldsymbol{r}(x, y) = (x, y, z(x, y))$

$$\boldsymbol{n}(x, y) = \frac{(x, y, \sqrt{R^2 - x^2 - y^2})}{R} = \frac{\boldsymbol{r}(x, y)}{r(x, y)} \equiv \frac{\boldsymbol{r}(x, y)}{R}. \tag{1a}$$

This surface appears to be convex from the point of view of incident beam. Similarly, for the bottom half-sphere $z(x, y) = -\sqrt{R^2 - x^2 - y^2}$ the unit vector normal to the concave surface is

$$\boldsymbol{n}(x, y) = \frac{(-x, -y, \sqrt{R^2 - x^2 - y^2})}{R} = \frac{-\boldsymbol{r}(x, y)}{r(x, y)} \equiv \frac{-\boldsymbol{r}(x, y)}{R}. \tag{1b}$$

Probably the formula

$$R - \sqrt{R^2 - x^2 - y^2} = R \frac{\dfrac{x^2 + y^2}{R^2}}{1 + \sqrt{1 - \dfrac{x^2 + y^2}{R^2}}} \tag{1c}$$

can be useful for numerical purposes when $x^2 + y^2 \ll R^2$.

## 2. Light beam incidenting the curved surface

Usually for the beam incidenting the curved surface we can define the central (mean or chief) ray that is for instance possible to identify with the symmetry axis of a wide beam of circular cross-section. For the converged beams with angle of convergence $\beta > 0$ or divergent beams with angle of divergence $\beta < 0$ or collimated beams ($\beta = 0$) we can define the propagation direction of the central ray by a unit vector $\mathbf{n}_{i,centr} = (n_{i,centr,x}, n_{i,centr,y}, n_{i,centr,z})$. Using standard spherical co-ordinates $\varphi_i \in \langle 0, 2\pi \rangle$, $\theta_i \in \langle \pi/2, \pi \rangle$ ($\cos\theta_i \leq 0$) this vector can be expressed as

$$\mathbf{n}_{i,centr} = (\cos\varphi_i \sin\theta_i, \sin\varphi_i \sin\theta_i, \cos\theta_i). \tag{2}$$

Let $\mathbf{r}_0 = (0, 0, R)$ is a point where the central beam incidents the sphere surface, then a circle with centre $\mathbf{r}_0$ and diameter $d$ in the plane perpendicular to $\mathbf{n}_{i,centr}$ is a set of points

$$C \equiv \{\mathbf{r}: \|\mathbf{r} - \mathbf{r}_0\| \leq d/2 \land \mathbf{n}_{i,centr} \cdot (\mathbf{r} - \mathbf{r}_0) = 0\}. \tag{3}$$

We may construct $C$ (that can be generalised for arbitrary beam cross sections) using two unit vectors $\mathbf{a}_1$, $\mathbf{a}_2$ composing the orthonormal base together with $\mathbf{n}_{i,centr}$ in the following way:

(a) When $\theta_i = \dfrac{\pi}{2}$ (or $n_{iz} = 0$, i.e. tangential incidence), then

$$\mathbf{a}_1 = (n_{i,centr,y}, -n_{i,centr,x}, 0) \equiv (\sin\varphi_i, -\cos\varphi_i, 0). \tag{4}$$

(b) When $\theta_i > \dfrac{\pi}{2}$, (i.e. $-1 \leq n_{i,centr,z} = \cos\theta_i < 0$), then for instance

$$\mathbf{a}_1 = (-n_{i,centr,z}, 0, n_{i,centr,x}) / \sqrt{1 - n_{i,centr,y}^2} \equiv (-\cos\theta_i, 0, \cos\varphi_i \sin\theta_i) / \sqrt{1 - (\sin\varphi_i \sin\theta_i)^2}. \tag{5}$$

Of course in these both cases the scalar product $\mathbf{a}_1 \cdot \mathbf{n}_{i,centr} = 0$. Especially $\varphi_i = \pi/2$ gives $\mathbf{n}_{i,centr} = (0, \sin\theta_i, \cos\theta_i)$ and $\mathbf{a}_1 = (1, 0, 0)$, since $-\cos\theta_i / \sqrt{1 - \sin^2\theta_i} = 1$. The other required vector $\mathbf{a}_2$ can be obtained as a cross product:

$$\mathbf{a}_2 = \mathbf{a}_1 \times \mathbf{n}_{i,centr}. \tag{6}$$

Especially we have

(a) for $\theta_i = \dfrac{\pi}{2}$, $\mathbf{a}_2 = (0, 0, 1)$,

(b) for $\theta_i > \dfrac{\pi}{2}$, $\mathbf{a}_2 = (-n_{i,centr,x} n_{i,centr,y}, 1 - n_{i,centr,y}^2, -n_{i,centr,y} n_{i,centr,z}) / \sqrt{1 - n_{i,centr,y}^2}$ and descriptive example $\varphi_i = \pi/2$ provides $\mathbf{a}_2 = (0, -\cos\theta_i, \sin\theta_i)$.

We can express an arbitrary element of $C$ using local Cartesian co-ordinates $(x', y')$ in the form of a sum of two vectors $\mathbf{r}_0 + \mathbf{a}(x', y')$, where

$$\mathbf{a}(x', y') = \mathbf{a}_1 x' + \mathbf{a}_2 y' \tag{7}$$

is a linear combination of the base vectors $\mathbf{a}_1$, $\mathbf{a}_2$ meeting the condition $x'^2 + y'^2 \leq d^2/4$. Another possibility for construction of $C$ employs polar co-ordinates $(\gamma, q)$ defined by identities $x' = \gamma\cos(q)$, $y' = \gamma\sin(q)$, where $0 \leq \gamma \leq d/2$ and $0 \leq q < 2\pi$.

The unit vector characterising the direction of the single ray propagation, however changing linearly (as postulated) with the distance $a = \|\mathbf{a}\|$, is then

$$\mathbf{n}_i = \frac{\dfrac{d}{2}\mathbf{n}_{i,centr} - \mathbf{a}\tan\left(\dfrac{\beta}{2}\right)}{\left\|\dfrac{d}{2}\mathbf{n}_{i,centr} - \mathbf{a}\tan\left(\dfrac{\beta}{2}\right)\right\|}, \tag{8}$$

other variances of this simple premise are also acceptable. For the denominator in (Eq. 8) we can find

$$\left\| \frac{d}{2} \boldsymbol{n}_{i,\text{centr}} - \boldsymbol{a} \tan\left(\frac{\beta}{2}\right) \right\| = \sqrt{\left(\frac{d}{2}\right)^2 + a^2 \tan^2 \frac{\beta}{2}}. \qquad (9)$$

A vector line equation with parameter $t_i \in (-\infty, +\infty)$ representing single ray of the incident beam is therefore:

$$\boldsymbol{r}_i(t_i) = \boldsymbol{r}_0 + \boldsymbol{a} + t_i \boldsymbol{n}_i \qquad (10)$$

and a first intersection of this ray with the reflecting surface derived from the quadratic equation $\|\boldsymbol{r}_i(t_i)\| = R$ is in detail

$$\boldsymbol{r}_i = \boldsymbol{r}_0 + \boldsymbol{a} + t_i \boldsymbol{n}_i, \qquad (11a)$$

$$t_i = -(Rn_{iz} + \boldsymbol{a} \cdot \boldsymbol{n}_i) - \sqrt{(Rn_{iz} + \boldsymbol{a} \cdot \boldsymbol{n}_i)^2 - (a^2 + 2Ra_z)}, \qquad (11b)$$

where index $z$ means $z$-component of the vector. Generally (except collimated beams) vectors $\boldsymbol{a}$, $\boldsymbol{n}_i$ are not mutually perpendicular. The discriminant (i.e. an expression under the root sign in Eq. 11b) of the solved quadratic equation has to be positive for existing ray intersection with the sphere and is just equal to zero for tangential incidence.

## 3. Polarisation of incident rays

The normal to the surface is $\boldsymbol{n} = \boldsymbol{r}_i / r_i$ ($\boldsymbol{r}_i$ is defined in Eqs. 11a, b) and therefore two local unit vectors, projecting two orthogonal components and allowing to define in this way the polarisation state of the electric fields, are

$$\boldsymbol{s} = \frac{\boldsymbol{n}_i \times \boldsymbol{n}}{\|\boldsymbol{n}_i \times \boldsymbol{n}\|}, \quad \boldsymbol{p} = \boldsymbol{s} \times \boldsymbol{n}_i. \qquad (12a)$$

To be noted that the polarisation state is not defined in the case of normal incidence when $\boldsymbol{n}_i = -\boldsymbol{n}$. Vectors $\boldsymbol{p}, \boldsymbol{s}, \boldsymbol{n}_i$ create right-handed orthonormal system, so s polarisation component of electric field is its projection to the vector $\boldsymbol{s}$, similarly for p polarisation component. In addition to the local quantities (Eq. 12a), varying with particular convergent or divergent rays, we can define similar useful quantities related to the central ray by formulae:

$$\boldsymbol{s}_{\text{centr}} = \frac{\boldsymbol{n}_{i,\text{centr}} \times \boldsymbol{n}_{\text{centr}}}{\|\boldsymbol{n}_{i,\text{centr}} \times \boldsymbol{n}_{\text{centr}}\|}, \quad \boldsymbol{p}_{\text{centr}} = \boldsymbol{s}_{\text{centr}} \times \boldsymbol{n}_{i,\text{centr}}, \text{ where } \boldsymbol{n}_{\text{centr}} = (0,0,1). \qquad (12b)$$

The polarisation state of the incident light, characterised by the complex electric field vector $\boldsymbol{E}_{i,\text{centr}} = (E_{ix}, E_{iy}, E_{iz})$, provided that it is the same for all beam rays (what can be certainly generalised to arbitrary beam polarisation profiles), can be determined just using vectors specified in (Eq. 12b). The complex variables

$$E_{i,\text{centr},s} = \boldsymbol{E}_{i,\text{centr}} \cdot \boldsymbol{s}_{\text{centr}}, \quad E_{i,\text{centr},p} = \boldsymbol{E}_{i,\text{centr}} \cdot \boldsymbol{p}_{\text{centr}} \qquad (13)$$

are s and p components of electric field of the central beam and presumably also of all rays of the incident beam, hence the electric field vector of the arbitrary ray is

$$\boldsymbol{E}_i = E_{i,\text{centr},s} \boldsymbol{s} + E_{i,\text{centr},p} \boldsymbol{p}. \qquad (14)$$

Actually, for each ray the s and p components of the field are the same as for the central ray in consistency with postulation

$$E_{i,s} \equiv \boldsymbol{E}_i \cdot \boldsymbol{s} = E_{i,\text{centr},s}, \quad E_{i,p} \equiv \boldsymbol{E}_i \cdot \boldsymbol{p} = E_{i,\text{centr},p}. \qquad (15)$$

The polarisation state of the incident beam can be also defined applying ellipsometric parameters $\psi_{\text{centr}}$, $\Delta_{\text{centr}}$ according to the formula

$$E_{i,p} / E_{i,s} \equiv E_{i,p,\text{centr}} / E_{i,s,\text{centr}} = \tan \psi_{\text{centr}} \exp(i\Delta_{\text{centr}}), \qquad (16a)$$

from which we have

$$E_{i,p} = E_0 \tan\psi_{centr} \exp(i\Delta_{centr}), \quad E_{i,s} = E_0, \tag{16b}$$

where $E_0$ is nonzero complex constant with absolute value proportional to the square root of incident light intensity.

It is obvious that instead of the vector base $a_1$, $a_2$ we can choose vectors $s_{centr}$, $p_{centr}$ satisfying the transformation identities

$$a_1 = \cos(q_0) s_{centr} + \sin(q_0) p_{centr}, \quad a_2 = -\sin(q_0) s_{centr} + \cos(q_0) p_{centr}, \tag{17a}$$

where

$$\cos(q_0) = a_1 \cdot s_{centr} = a_2 \cdot p_{centr}, \quad \sin(q_0) = a_1 \cdot p_{centr} = -a_2 \cdot s_{centr}. \tag{17b}$$

## 4. Polarisation of the reflected beam

Unit vectors defining the directions of the locally reflected rays can be obtained from the equation

$$n_r = n_i - 2(n_i \cdot n)n \tag{17}$$

and local angles of incidence $\alpha_i$ (identical to the angles of reflection $\alpha_r$) are

$$\alpha_i \equiv \alpha_r = \arccos(n \cdot n_r) = -\arccos(n \cdot n_i). \tag{18}$$

Designating the local amplitude reflectivities of the curved surface for s and p electromagnetic waves by $r_s$, $r_p$ (they are complex numbers depending on the angle of incidence, wavelength and optical properties of reflecting material) we can calculate the amplitudes of reflected electric field for each polarisation state of the ray

$$E_{r,s} = r_s E_{i,s}, \quad E_{r,p} = r_p E_{i,p}, \tag{19}$$

so that the electric field vector of the reflected ray is

$$E_r = E_{r,s} s + E_{r,s} p \tag{20}$$

taking into account that unit vector $s$ (normal to the plane of incidence) and unit vector $p$ (lying in the plane of incidence) are both identical for the ray incidenting the surface and ray reflected from the surface. The correct ellipsometric parameters $\psi_{corr}$, $\Delta_{corr}$ of the reflected wave are then defined from the relation

$$\tan\psi_{corr} \exp(i\Delta_{corr}) = r_p / r_s, \tag{21}$$

hence

$$\tan\psi_{corr} \exp(i\Delta_{corr}) = \frac{E_{r,p}}{E_{r,s}} \frac{E_{i,s}}{E_{i,p}} = \frac{E_{r,p}}{E_{r,s}} \frac{1}{\tan\psi_{centr} \exp(i\Delta_{centr})}. \tag{22}$$

Using (Eq. 22) it is possible to solve the inverse ellipsometric problem, for instance for single dielectric layer deposited on the substrate of known refraction index, for all rays incidenting the curved surface under different local angles of incidence, and then to obtain the same index of refraction and thickness of the layer, although evidently the input ellipsometric parameters locally differ.

Otherwise, in the case when a light detector (for instance ideal rotating analysator capable to distinguish and analyse single beam rays with sufficient space resolution) is calibrated for the central ray (see definition of vectors $s_{centr}$ and $p_{centr}$ by Eq. 12b), we obtain (intentionally) inaccurate ellipsometric parameters

$$\tan\psi \exp(i\Delta) = \frac{(E_r \cdot p_{centr})}{(E_r \cdot s_{centr})} \frac{1}{\tan\psi_{centr} \exp(i\Delta_{centr})}. \tag{23}$$

It is clear that quantities $\psi$ a $\Delta$ from (Eq. 23) hold true precisely only for the central ray, however for all other rays of the light beam allow to determine quantitatively the deviations from the correct values and therefore to estimate the error emerged from measurement ellipsometric parameters of curved surfaces in comparison with a flat surface and also to analyse the effect of beam convergence or divergence in comparison with collimated beam.

## 5. Inverse ellipsometric problem for a single thin dielectric layer

The aim of this section is to find the thickness and index of refraction of single dielectric layer from known ellipsometric parameters $\psi$ and $\Delta$. For this purpose we define quantities $n_0$ – index of refraction of external environment (superstrate, air), $n_1$ – (unknown) real index of refraction of the layer with (unknown) thickness $h_1$, $n_g$ – index of refraction of the substrate (generally complex number), $\alpha_0$ – angle of incidence measured in the superstrate and $\lambda$ – wavelength of light in vacuum. We also use the admittance of vacuum $y_0 = \sqrt{\varepsilon_0/\mu_0}$ ($\mu_0$, $\varepsilon_0$ – permeability and permittivity of vacuum) for definition of relative admittances for both light polarisations of all media considering the law of refraction in the form $n_0 \sin\alpha_0 = n_1 \sin\alpha_1 = n_g \sin\alpha_g$, see for instance [1]

$$Y_0^{(s)} = \sqrt{n_0^2 - (n_0 \sin\alpha_0)^2} = n_0 \cos\alpha_0, \quad Y_0^{(p)} = n_0^2/\sqrt{n_0^2 - (n_0 \sin\alpha_0)^2} = n_0/\cos\alpha_0,$$

$$Y_1^{(s)} = \sqrt{n_1^2 - (n_0 \sin\alpha_0)^2} = n_1 \cos\alpha_1, \quad Y_1^{(p)} = n_1^2/\sqrt{n_1^2 - (n_0 \sin\alpha_0)^2} = n_1/\cos\alpha_1, \qquad (24)$$

$$Y_g^{(s)} = \sqrt{n_g^2 - (n_0 \sin\alpha_0)^2} = n_g \cos\alpha_g, \quad Y_g^{(p)} = n_g^2/\sqrt{n_g^2 - (n_0 \sin\alpha_0)^2} = n_g/\cos\alpha_g.$$

The phase shift of plane wave propagating once through the layer $\varphi_1$ is independent on the polarisation and is equal to

$$\varphi_1 = \frac{2\pi}{\lambda} h_1 \sqrt{n_1^2 - (n_0 \sin\alpha_0)^2}. \qquad (25)$$

The solution of Maxwell equations for plane monochromatic waves in isotropic medium can be expressed by $2\times 2$ matrix $\mathbf{S}^{(s,p)}$ physically representing transform operators for tangential components of electric field vectors, however separated into counter-propagating waves

$$\mathbf{S}^{(s,p)} = \frac{1}{2}\begin{pmatrix} 1 & \frac{1}{Y_0^{(s,p)}} \\ 1 & -\frac{1}{Y_0^{(s,p)}} \end{pmatrix}\begin{pmatrix} \cos\varphi_1 & \frac{i}{Y_1^{(s,p)}}\sin\varphi_1 \\ iY_1^{(s,p)}\sin\varphi_1 & \cos\varphi_1 \end{pmatrix}\begin{pmatrix} 1 & 1 \\ Y_g^{(s,p)} & -Y_g^{(s,p)} \end{pmatrix} \qquad (26)$$

for both polarisation states. From matrix elements of $\mathbf{S}^{(s,p)}$ we can compute eight amplitude coefficients, namely amplitude reflectivities for waves propagating in the direction from superstrate to the substrate in the form

$$r_{s,p} = \frac{S_{21}^{(s,p)}}{S_{11}^{(s,p)}}. \qquad (27)$$

Employing the definition of ellipsometric parameters

$$\tan\psi \exp(i\Delta) = \frac{r_p}{r_s}, \qquad (28)$$

considered in this moment known (for instance as a result of measurement) we can derive from (Eqs. 27, 28) the quadratic equation - a condition for the wanted layer parameters [1]

$$a_0 \exp(-4i\varphi_1) + b_0 \exp(-2i\varphi_1) + c_0 = 0. \qquad (29)$$

Here

$$a_0 = b^{(s)}d^{(p)}\tan\psi\exp(i\Delta) - d^{(s)}b^{(p)},$$

$$b_0 = (a^{(s)}d^{(p)} + b^{(s)}c^{(p)})\tan\psi\exp(i\Delta) - (d^{(s)}a^{(p)} + c^{(s)}b^{(p)}), \tag{30}$$

$$c_0 = a^{(s)}c^{(p)}\tan\psi\exp(i\Delta) - c^{(s)}a^{(p)},$$

where quantities with s or p indices (not all are independent)

$$a^{(s,p)} = (1 - Y_1^{(s,p)}/Y_0^{(s,p)})(1 + Y_g^{(s,p)}/Y_1^{(s,p)}), \quad b^{(s,p)} = (1 + Y_1^{(s,p)}/Y_0^{(s,p)})(1 - Y_g^{(s,p)}/Y_1^{(s,p)}), \tag{31}$$

$$c^{(s,p)} = (1 + Y_1^{(s,p)}/Y_0^{(s,p)})(1 + Y_g^{(s,p)}/Y_1^{(s,p)}), \quad d^{(s,p)} = (1 - Y_1^{(s,p)}/Y_0^{(s,p)})(1 - Y_g^{(s,p)}/Y_1^{(s,p)}).$$

The solution of (Eq. 29) has to be in the form of complex units (due to the fact that $\varphi_1$ is real), therefore the complex conjugated equation

$$c_0^*\exp(-4i\varphi_1) + b_0^*\exp(-2i\varphi_1) + a_0^* = 0, \tag{32}$$

where asterisk means complex conjugated quantities, have to be fulfilled too. Therefore we can reduce the quadratic equation for $\exp(-2i\varphi_1)$ to the linear equation and a condition for index of refraction

$$f(n_1) \equiv \left|b_0^*c_0 - b_0 a_0^*\right|^2 - \left|\left|a_0\right|^2 - \left|c_0\right|^2\right|^2 = 0. \tag{33}$$

Admittable layer thicknesses are then

$$h_k = \left(k - \frac{\sigma}{2\pi}\right)h_{\text{per}}, \quad k = 1, 2, \ldots, \tag{34}$$

where

$$\exp(i\sigma) = \frac{b_0^*c_0 - b_0 a_0^*}{|a_0|^2 - |c_0|^2}, \quad 0 \le \sigma < 2\pi, \quad h_{\text{per}} = \frac{\lambda}{2\sqrt{n_1^2 - (n_0\sin\alpha_0)^2}}, \tag{35}$$

$h_{\text{per}}$ is a period of the layer thickness. This method is more efficient than least square method applied for two unknown variables at the same time. From local ellipsometric variables $\psi_{\text{corr}}$, $\Delta_{\text{corr}}$ (Eq. 22) and using the above relations we can compute a local layer parameters identical with the given layer index of refraction and its thickness because these quantities have to be independent on the ray selection. Certainly, we have different values when computing the ellipsometric parameters obtained from the detector calibrated for the central beam (Eq. 23), which are intentionally affected by the systematic error due to the measuring method.

## 6. Averaging of ellipsometric parameters

To average values of quantities over beam cross-section we can use an equidistant grid so that the electric field of each ray is considered with the same weight what allows to approximate integrals over surface sufficiently. For this purpose the following equidistant square grid spread over $C$ may be suitable

$$x_k' = \frac{d}{2}\left(-1 + 2\frac{k-1}{N-1}\right), \quad k = 1, \ldots, N, \tag{36a}$$

$$y_l' = \frac{d}{2}\left(-1 + 2\frac{l-1}{N-1}\right), \quad l = 1, \ldots, N. \tag{36b}$$

But only the subset $M$ of the ordered pairs of natural numbers refers actually to the beam, namely in the case of the circular beam cross-section it is a set

$$M = \left\{(k,l) \in <1,N> \times <1,N> : x_k'^2 + y_l'^2 \le \left(\frac{d}{2}\right)^2\right\}. \tag{37}$$

The points $r_0 + a_1 x'_k + a_2 y'_l$, $(k,l) \in M$ (their number is $m(M)$) may be used for calculation of averaged variables expecting the more precise result for larger $N$. There are at least three ways of averaging:

(a) Averaging local index of refraction $n_1(k,l)$ and thickness $h_1(k,l)$ following the relations

$$n_{1,\text{aver}}^{(a)} = \frac{1}{m(M)} \sum_{(k,l) \in M} n_1(k,l), \quad h_{1,\text{aver}}^{(a)} = \frac{1}{m(M)} \sum_{(k,l) \in M} h_1(k,l). \quad (38)$$

(b) Averaging local ellipsometric parameters obtained via (Eq. 23)

$$\psi_{\text{aver}}^{(b)} = \frac{1}{m(M)} \sum_{(k,l) \in M} \psi(k,l), \quad \Delta_{\text{aver}}^{(b)} = \frac{1}{m(M)} \sum_{(k,l) \in M} \Delta(k,l) \quad (39)$$

and then applying the algorithm explained in the section 5 for calculation averaged quantities $n_{1,\text{aver}}^{(b)}$ and $h_{1,\text{aver}}^{(b)}$.

(c) Averaging the electric field vector of the reflected beam (Eq. 20)

$$\boldsymbol{E}_{\text{r,aver}} = \frac{1}{m(M)} \sum_{(k,l) \in M} \boldsymbol{E}_r(k,l), \quad (40)$$

then using (Eq. 23) in the form

$$\tan \psi \exp(i\Delta) = \frac{(\boldsymbol{E}_{\text{r,aver}} \cdot \boldsymbol{p}_{\text{centr}})}{(\boldsymbol{E}_{\text{r,aver}} \cdot \boldsymbol{s}_{\text{centr}})} \frac{1}{\tan \psi_{\text{centr}} \exp(i\Delta_{\text{centr}})} \quad (41)$$

and finally calculating $n_{1,\text{aver}}^{(c)}$ and $h_{1,\text{aver}}^{(c)}$ from such obtained ellipsometric parameters.

## 7. Numerical simulation

The following example demonstrates the effect of the curved surface and convergence or divergence of the incident beam on the evaluation of ellipsometric data: the spherically shaped silicon substrate with radius $R = 1$ m and a complex index of refraction $n_g = 3.855 - 0.024\text{i}$ is covered by a thin layer with index of refraction $n_1 = 1.461$ (silicon oxide) and thickness $h_1 = 112$ nm, the index of refraction of superstrate is $n_0 = 1$ (air). The circular beam with diameter $d$ (variable quantity) and wavelength $\lambda = 632.8$ nm incidents the sphere from the direction $\varphi_i = 90°$, $\theta_i = 110°$, it means that the angle of incidence measured in the superstrate is $\alpha_0 \equiv \alpha_i = 70°$ and a central beam is parallel to the plane $x = 0$. Considered ellipsometric parameters of the linearly polarised incident beam are $\psi_{\text{centr}} = 45°$, $\Delta_{\text{centr}} = 0°$ and a number of grid division $N = 150$.

Figures illustrating maps of local quantities are depicted for the diameter of the collimated beam $d = 2R(1 - \sin \alpha_i) \approx 120.6$ mm, it is an extreme case when just one boundary ray incidents the sphere tangentially. It is clear that the problem exhibits the dimensionless characteristic number $d/R$.

The change of the local angle of incidence of the collimated beam ($\beta = 0$) could be seen in *Fig. 1*, the values vary between $61,57°$ and $90°$.

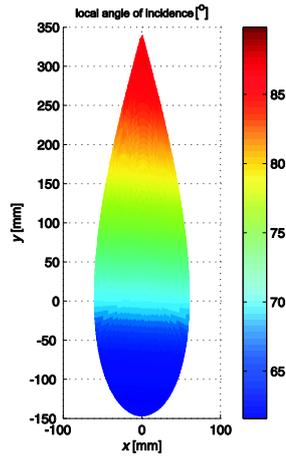

Fig. 1 Local angle of incidence of the collimated beam with the diameter 120.6 mm impinging the sphere with radius 1 m; angle of incidence of the central beam is $70^\circ$.

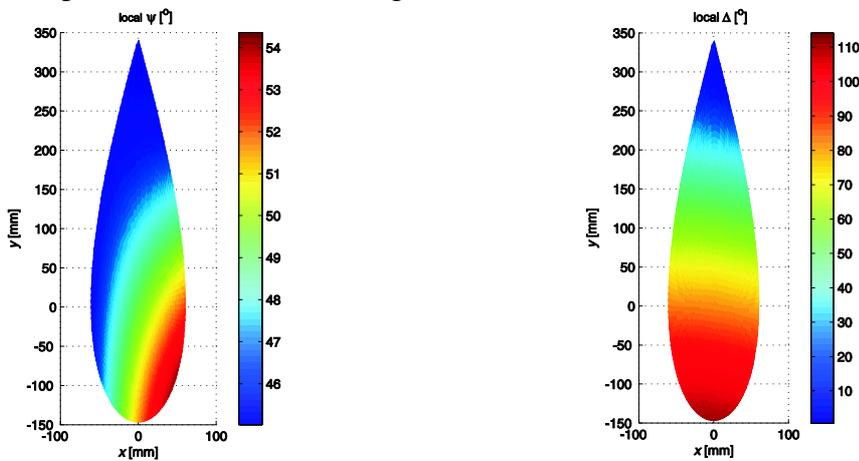

Fig. 2 Local ellipsometric parameters $\psi$ (left) $\Delta$ (right) for same conditions as in Fig. 1.

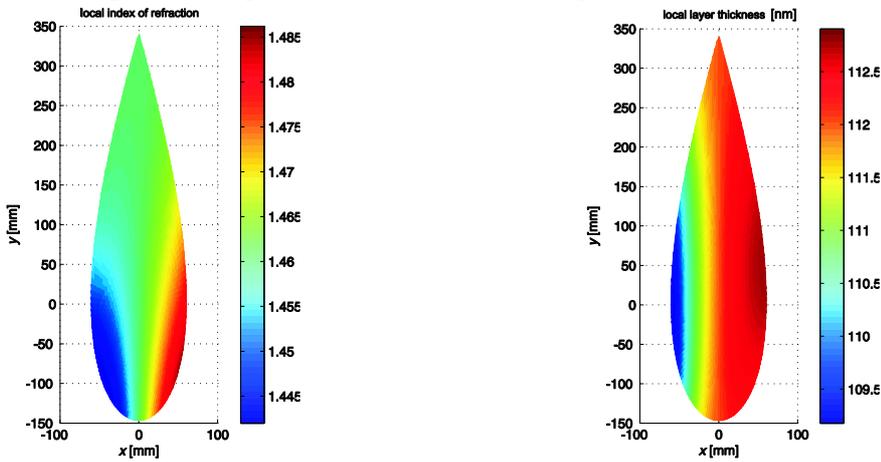

Fig. 3 Local index of refraction (left) and thickness (right) of the layer with (intentional) error for same conditions as in Fig. 1.

*Fig. 2* demonstrates local changes of ellipsometric parameters. Using local angles of incidence it is possible to calculate the correct layer index of refraction and its thickness. The *Fig. 3* shows local erroneous values of layer index of refraction and its thickness obtained from ellipsometric parameters according to (Eq. 23) when the local polarisation state of the each beam is replaced by the polarisation state of the central beam.

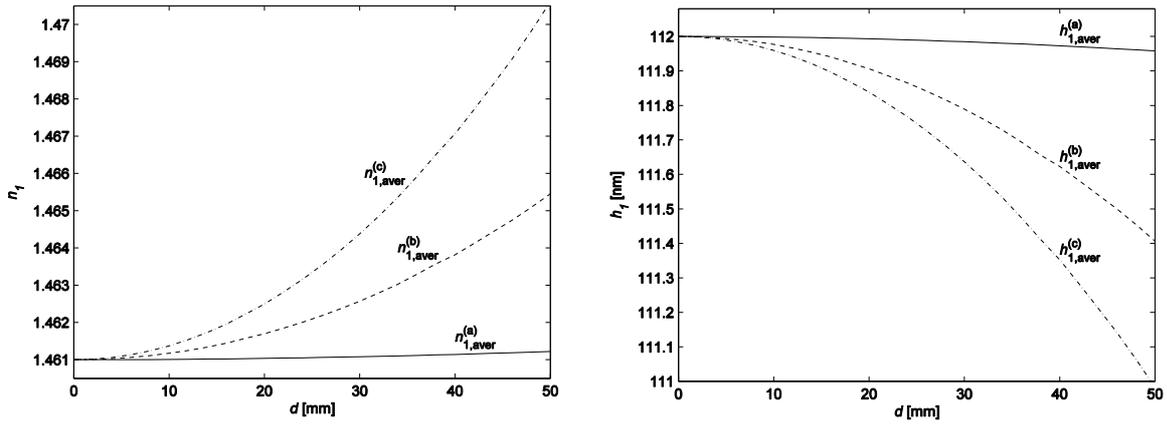

Fig. 4 Averaged values of the layer index of refraction (left) and its thickness (right) obtained by three methods in dependence on collimated beam diameter $d$, the radius of the sphere is $R = 1$ m.

From *Fig. 4* we can deduce the error rate for determination of thin layer parameters with increasing diameter of the collimated beam. The more precise seems to be (a) method in comparison with (b) method, the result is nearly constant for $d/R$ roughly less than 5 %. The more adverse method is (c) using for detection the resulting electric field of the reflected beam, but it is just a method supposed to approximate the physical measurement conditions more realistically. However, also for relatively wide beam with diameter 50 mm we obtain the deviation only 0.01 from the correct index of refraction 1.461 and for layer thickness the value 111 nm while the correct value is 112 nm.

The effect of beam convergence or divergence on the measurement precision is demonstrated in *Figs. 5 – 8*.

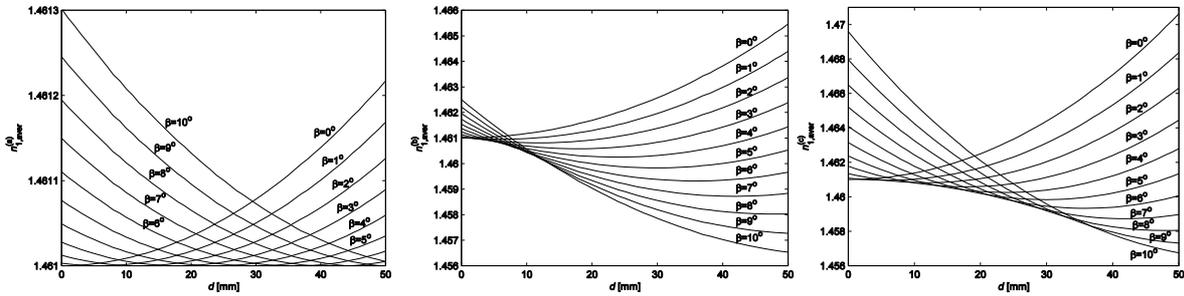

Fig. 5 Averaged index of refraction of thin layer obtained by three averaging methods (a, b, c from left to right) from measurement of ellipsometric parameters for convergent beam with angle of convergence $\beta = 0° - 10°$.

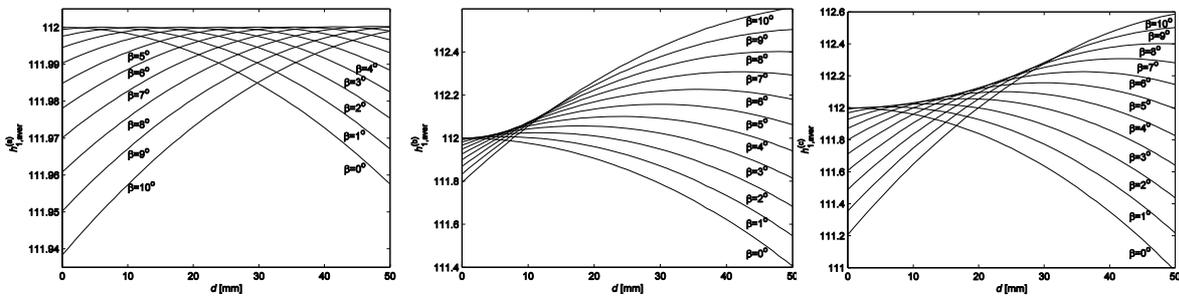

Fig. 6 Averaged thickness of thin layer obtained by three averaging methods (a, b, c from left to right) from measurement of ellipsometric parameters for convergent beam with angle of convergence $\beta = 0° - 10°$.

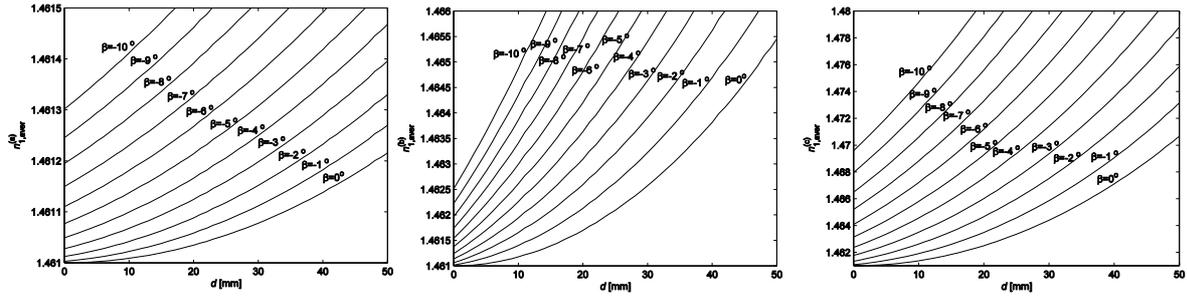

Fig. 7 Averaged index of refraction of thin layer obtained by three averaging methods (a, b, c from left to right) from measurement of ellipsometric parameters for divergent beam with angle of divergence $\beta = -10^\circ - 0^\circ$.

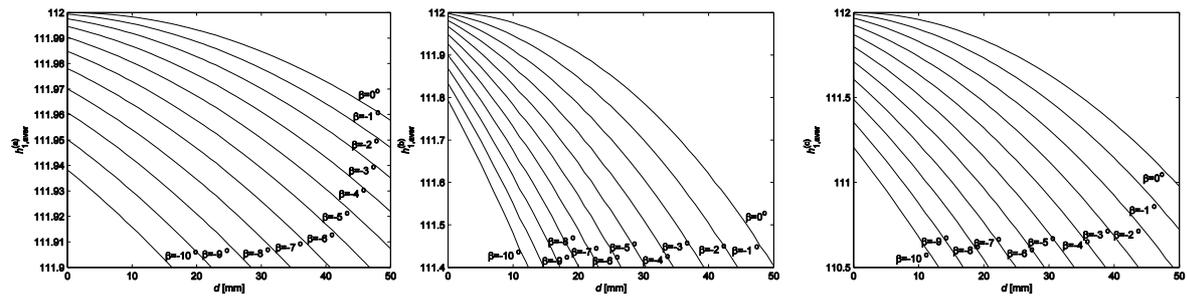

Fig. 8 Averaged thickness of thin layer obtained by three averaging methods (a, b, c from left to right) from measurement of ellipsometric parameters for divergent beam with angle of divergence $\beta = -10^\circ - 0^\circ$.

These figures allow to estimate quantitatively the effect of beam convergence or divergence on the inaccuracy of results obtained from ellipsometric measurements. Surprisingly we can expect better results for convergent beams in comparison with collimated beams but only for suitable diameter of the beam. In the simulated case the smallest deviations of the index of refraction and thickness from the correct values occur when the convergent angle is about five degrees. Divergent beams unambiguously exhibit worse results for all angles of beam divergence with increasing beam diameter and angle of divergence. The averaging method using the local index of refraction and thickness obtained from locally measured ellipsometric parameters (i.e. method (a)) is minimally sensitive to the beam diameter and its degree of convergence or divergence.

## 8. Conclusion

This paper using the combination of geometrical and wave optics quantitatively examines the effect of curved surface and also beam convergence or divergence on the accuracy of the results obtained from ellipsometric measurements. For this purpose three averaging methods for estimation the measurable quantities were proposed and results numerically compared applying the example of one thin dielectric layer deposited on the spherically shaped substrate.

**Acknowledgement**
This work was supported by the project No. 1M06002 of the Ministry of Education, Youth and Sports of the Czech Republic.